\documentclass[prl,reprint,superscriptaddress,nofootinbib,twocolumn]{revtex4-1}

\usepackage[linktoc=page,colorlinks,urlcolor=blue,citecolor=blue,linkcolor=blue]{hyperref}
\usepackage{amssymb,amsmath}
\usepackage{graphicx}
\usepackage{color}
\usepackage{upgreek}
\usepackage{overpic}
\usepackage{gensymb}

\newcommand{\heFour}{$^{4}$He~}
\newcommand{\nFifteen}{$^{15}$N~}
\providecommand{\e}[1]{\ensuremath{\times 10^{#1}}}

\newcommand{\snl}{Sandia National Laboratories, Albuquerque, New Mexico 87185, USA}
\newcommand{\cint}{Center for Integrated Nanotechnologies, Sandia National Laboratories, Albuquerque, New Mexico 87123, USA}

\begin{document}
\title{A fitting algorithm for optimizing ion implantation energies and fluences}
\date{\today}
\author{Pauli Kehayias}
\email{pmkehay@sandia.gov}
\affiliation{\snl}
\author{Jacob Henshaw}
\affiliation{\cint}
\author{Maziar Saleh Ziabari}
\affiliation{\cint}
\affiliation{University of New Mexico Department of Physics and Astronomy, Albuquerque, New Mexico 87131, USA}
\author{Michael Titze}
\affiliation{\snl}
\author{Edward Bielejec}
\affiliation{\snl}
\author{Michael P. Lilly}
\affiliation{\cint}
\author{Andrew M. Mounce}
\affiliation{\cint}

\begin{abstract}
We describe a method to automatically generate an ion implantation recipe, a set of energies and fluences, to produce a desired defect density profile in a solid using the fewest required energies. We simulate defect density profiles for a range of ion energies, fit them with an appropriate function, and interpolate to yield defect density profiles at arbitrary ion energies.  Given $N$ energies, we then optimize a set of $N$ energy-fluence pairs to match a given target defect density profile. Finally, we find the minimum $N$ such that the error between the target defect density profile and the defect density profile generated by the $N$ energy-fluence pairs is less than a given threshold. Inspired by quantum sensing applications with nitrogen-vacancy centers in diamond, we apply our technique to calculate optimal ion implantation recipes to create uniform-density 1 $\upmu$m surface layers  of \nFifteen or vacancies (using $^4$He).
\end{abstract}
\maketitle

\section{Introduction}
Ion implantation and irradiation are critical techniques with multiple application areas.  These include the formation of dopant layers in semiconductor devices \cite{SiVLSIFundamentals} and the creation of color centers in diamond and other wide-bandgap materials for quantum optics and sensing, electron and hole dopants to alter the conductivity, and graphitization to make electrical contacts and membranes \cite{rabeauN15, NVCVDreview, kalishReview, tenBoronLayer, ohmicGraphite, evelynHuMembranes}.  Other examples include making uniform damage layers to alter the superconducting critical temperature $T_c$ in YBaCuO films \cite{YBaCuOTc} and to enable high-resolution diamond cutting \cite{highResDiamondCutting}, to name a few.  While knowing the implanted ion range and straggle for a given energy is important in these highlighted examples, there are many instances where a single-energy ion implantation cannot produce the desired ion or vacancy distribution in a material, including in the examples referenced above.  For these types of applications, the community typically uses Stopping Range of Ions in Matter (SRIM) simulations to predict the ion density and vacancy density depth distributions for ion implantations into solids \cite{SRIM2010}. However, manually calculating and choosing a set of implantation energies and fluences to satisfactorily match a desired defect density profile can be laborious and inaccurate, especially as the number of required implants increases.

In this paper, we describe a method to computationally generate ion implantation recipes for arbitrary defect density profiles using least-squares curve fitting. We do this by fitting a set of simulated defect density profiles from SRIM, interpolating the resulting fit parameters, and minimizing the number of ion implantation energy-fluence pairs required to produce a recipe that closely matches the desired defect density profile. 

We are motivated by the past success using SRIM simulations to calculate the ion implantation recipes to make shallow layers of nitrogen-vacancy (NV) defect centers in diamond, which are used for sensing magnetic sources external to the diamond surface \cite{NVnanogratings, hemozoin, victorHeImplant}. An NV center consists of a substitutional nitrogen atom in the diamond lattice adjacent to a vacancy. 

To create a surface layer of NV centers, one can implant nitrogen into a diamond sample with few-ppb impurity density (which also creates vacancies) or implant another ion (such as helium) into a diamond sample with nitrogen defects ($\sim$100 ppm) to create vacancies. The resulting NV layer distribution is expected to match that of the implanted nitrogen or vacancy density. Since the magnetic field amplitude that an NV center senses depends on the distance to the magnetic source (such as a magnetic dipole), tailoring the NV layer thickness and depth is important to optimize the magnetic signal-to-noise ratio. By predicting the implantation characteristics using SRIM, we engineer an ideal NV layer to suit our magnetometry specifications.

\begin{figure*}[ht]
\begin{center}
\begin{overpic}[width=0.98\textwidth]{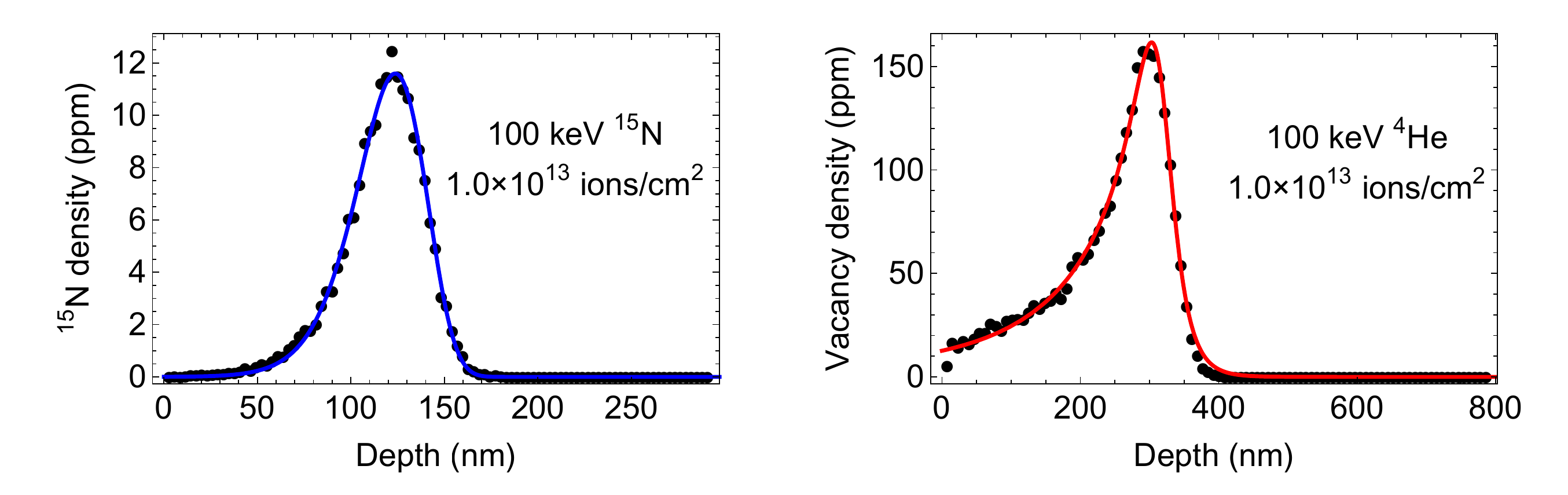}
\put(42.5,26.5){\textsf{\Large a}}
\put(92,26.5){\textsf{\Large b}}
\end{overpic}
\caption{\label{exampleSRIM}
(a) Simulated \nFifteen density $\rho_{\textrm{sim}}(z)$ (black) and $\rho_{\textrm{Gaus}}(z)$ fit (blue). 
(b) Simulated $^4$He-induced vacancy density $\rho_{\textrm{sim}}(z)$ (black) and $\rho_{\textrm{Lor}}(z)$ fit (red). The vacancy density is the sum of the vacancies created by the \heFour ions and the vacancies created by carbon recoils. For diamond, 1 ppm = 1.76\e{17} atoms/cm$^{3}$. The systematic residuals are small ($<$1\% of the maximum) compared to the statistical residuals (about 2\% of the maximum). 
}
\end{center}
\end{figure*}

\section{Methods}
Our ion implantation recipe optimization algorithm consists of the following steps:

\begin{enumerate}
  \item Simulate ion implantation defect density profiles for a list of ion energies using SRIM.
  \item Fit the resulting defect density profile using an empirical fit function, then interpolate the fit parameters as a function of energy.
  \item For a given number of ion implantation energies $N$, calculate the ion implantation recipe (energy-fluence pairs) that best fits to the target defect density profile.
  \item Repeat step 3 with increasing $N$ until the ion implantation recipe produces a defect density profile that resembles the target defect density profile within a given error limit.
\end{enumerate}

\noindent
We applied this algorithm to two situations relevant to creating shallow NV layers in diamond for magnetic sensing and imaging of near-surface external magnetic sources \cite{rabeauN15, NVCVDreview}. In the first example, we calculated the implantation recipe needed to create a flat-top \nFifteen density in a diamond substrate (few-ppb impurity density) with a 1 $\upmu$m thickness (similar to Sample B in Ref.~\cite{micromagnetPUFs}). In the second example, we calculated the \heFour implant recipe needed to create a 1 $\upmu$m uniform layer of vacancies in a diamond substrate. In this recipe, the diamond substrate already has an appreciable nitrogen density in the bulk ($\sim$100 ppm), and we implant with \heFour to create vacancies and convert nitrogen defects to NV defects \cite{hemozoin, victorHeImplant}.

\begin{figure*}
  \centering
  \includegraphics[width=0.98\textwidth]{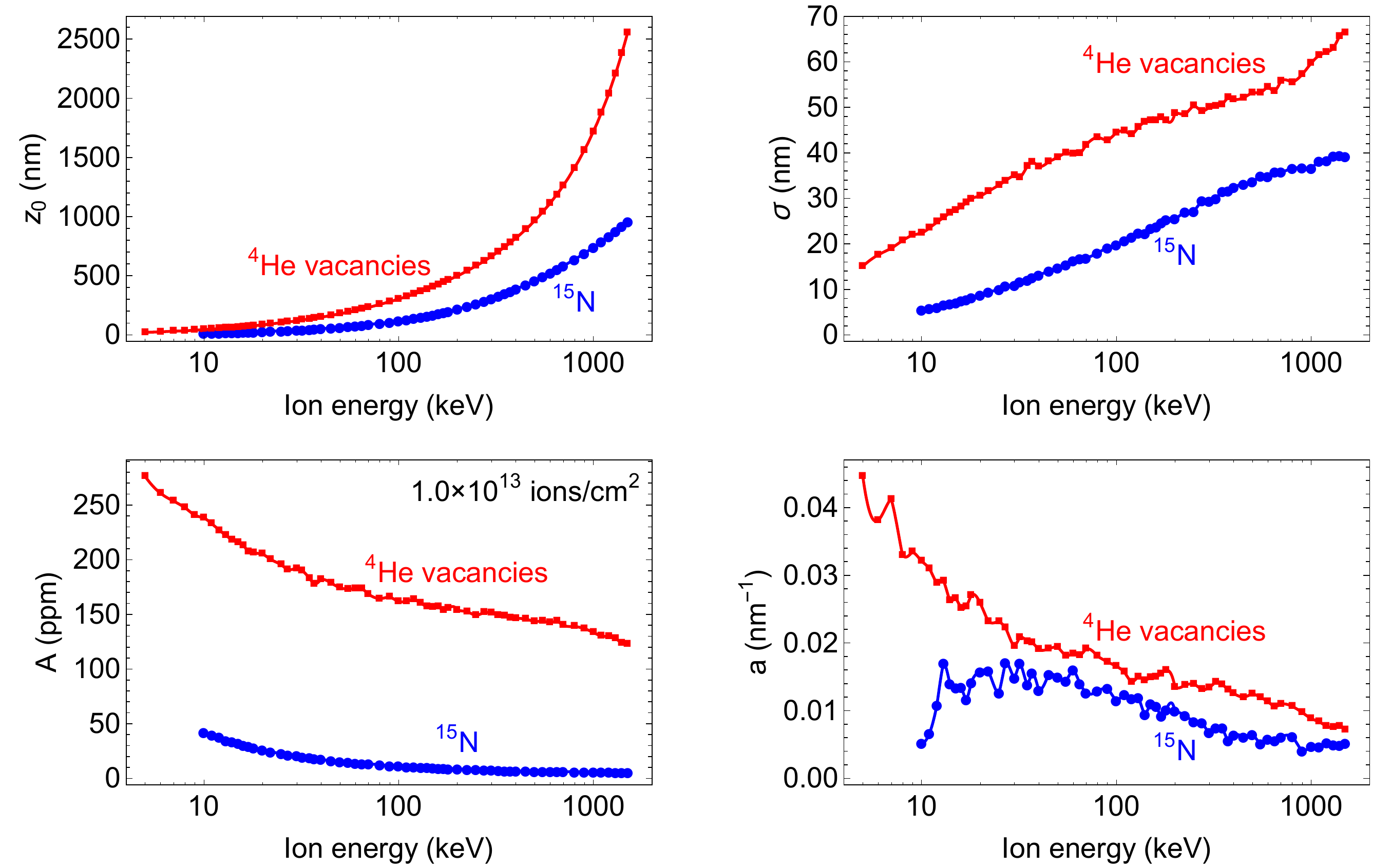}
  \caption{Extracted $z_0$, $\sigma$, $A$, and $a$ parameters for the \nFifteen asymmetric Gaussian fits (blue) and \heFour vacancy asymmetric Lorentzian fits (red). Here we assume a 1.0\e{13}/cm$^2$ fluence for each. }
  \label{allFitParams}
\end{figure*}

\subsection{Ion implantation and defect density profile simulations}
We simulated \nFifteen and \heFour ion implants using the SRIM Monte Carlo simulation software and the \texttt{pysrim} Python library \cite{SRIM2010, pysrim}. For each ion, we first generated the projected range and longitudinal straggle as a function of energy. This enabled us to assign an appropriate depth range for the transport of ions in matter (TRIM) simulation, which produced the defect density profiles used in later steps. For \nFifteen we simulated 58 energies spanning from 10 keV to 1500 keV (15 nm to 960 nm in  depth), and for \heFour we simulated 63 energies spanning from 5 keV to 1500 keV (22 nm to 2550 nm in  depth).

For each energy, we simulated 10,000 ions incident at 8$\degree$ from the diamond surface normal (usually the [100] crystallographic axis) to avoid channeling along  crystallographic axes in real diamond samples. The diamond substrate was modeled as a $^{12}$C solid with 3.51 g/cm$^3$ density, 37.5 eV atom displacement threshold energy, 7.35 eV lattice damage threshold, and 7.5 eV surface damage threshold \cite{NVnanogratings, diamondDisplEnergy}. Each simulation calculated the ion and vacancy densities for 100 evenly-spaced depth bins starting from the surface, and we expanded the depth spacing (and overall depth) with increasing ion energy.  Figure \ref{exampleSRIM} shows example defect density profiles for \nFifteen and \heFour implants (100 keV, 1.0\e{13} ions/cm$^2$).

\subsection{Defect density profile fitting}
For each implantation energy, we fit the simulated defect density profile $\rho_{\textrm{sim}}(z)$ as a function of depth $z$ using an asymmetric Gaussian (Lorentzian) distribution for ions (vacancies), defined as \cite{asymLorGaus}

\begin{equation}
\rho_{\textrm{Gaus}}(z) =  \frac{A(1 + e^{a(z-z_0)} )^2}{4} \exp \left[ - \frac{(1 + e^{a(z-z_0)} )^2}{4} \frac{(z-z_0)^2}{2 \sigma^2} \right],
\end{equation}
\begin{equation}
\rho_{\textrm{Lor}}(z) =  \frac{A \sigma^2}{\sigma^2 + \frac{(1 + e^{a(z-z_0)} )^2}{4} (z-z_0)^2}.
\end{equation}
\noindent
In these expressions, $z_0$ is a depth with maximum defect density, $\sigma$ is a linewidth, $A$ is a maximum amplitude,  and $a$ is an asymmetry parameter ($a=0$ corresponds to no asymmetry).  Figure \ref{exampleSRIM} shows the asymmetric Gaussian and Lorentzian fits used for two example $\rho_{\textrm{sim}}(z)$ distributions; while these fit functions are empirical, they are able to capture the simulated lineshapes well. The fit functions smooth out the statistical noise in $\rho_{\textrm{sim}}(z)$ and generate a continuous function for each defect density profile. After fitting each $\rho_{\textrm{sim}}(z)$ for our simulation range, we interpolate $z_0$, $\sigma$, $A$, and $a$ as a function of energy (Fig.~\ref{allFitParams}) to yield the defect density profile for any energy within the simulated range.

\begin{figure}[ht]
\begin{center}
\begin{overpic}[width=0.48\textwidth]{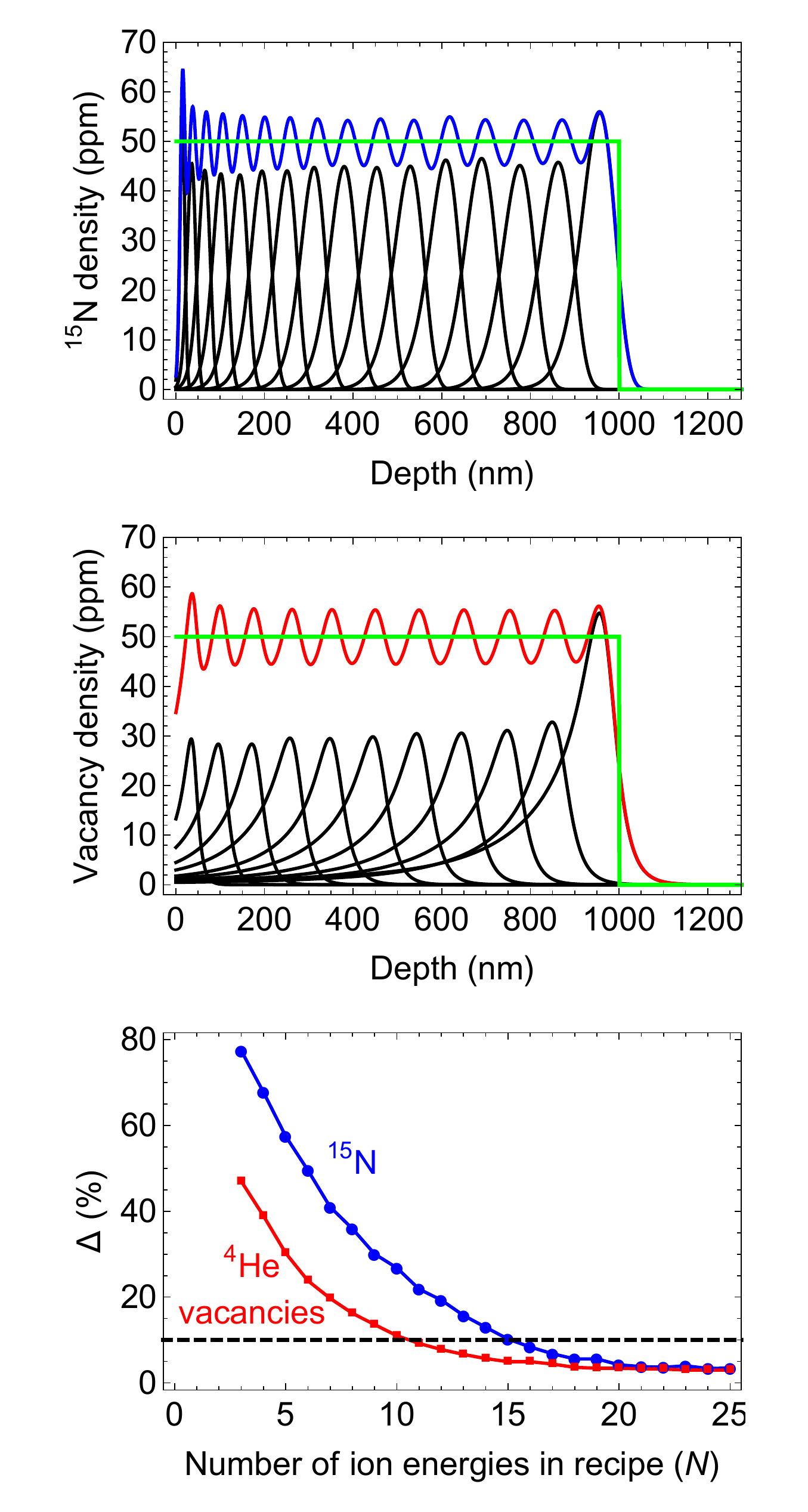}
\put(46,94){\textsf{\Large a}}
\put(46,60.5){\textsf{\Large b}}
\put(46,28.5){\textsf{\Large c}}
\end{overpic}
\caption{\label{exampleDoses}
Fitted $\rho_{\textrm{recipe}}(z,N)$ defect density profiles for a 1 $\upmu$m 50 ppm flat-top for \nFifteen (a) and $^4$He-induced vacancies (b). The black curves represent the defect density profiles from the constituent implants, which sum to approximate a uniform flat-top density $\rho_{\textrm{target}}(z)$ (green). (c) Percent error between $\rho_{\textrm{target}}(z)$ and $\rho_{\textrm{recipe}}(z,N)$ for an increasing number of energies $N$. The dashed line indicates $\Delta_{\textrm{th}} = 10$\%, which is the threshold we use for this work. 
}
\end{center}
\end{figure}

\subsection{Ion implantation energy and fluence optimization}
For a set of $N$ energies and fluences in an ion implantation recipe, the total defect density profile is 

\begin{equation}
  \rho_{\textrm{recipe}}(z,N) =  \sum_{i=1}^{N} f_i \rho_{E_i}(z)~. 
\end{equation}
\noindent
Here, $E_i$ is the energy for the $i$'th ion implant, $f_i$ is the fluence, and $\rho_{E_i}(z)$ is the defect density profile for the $i$'th ion implant and energy ($\rho_{\textrm{Gaus}}(z)$ or $\rho_{\textrm{Lor}}(z)$ with interpolated $\{z_0, \sigma, A, a\}$ parameters). For a given $N$, we perform a least-squares fit, minimizing $\int_{0}^{\infty} \left[ \rho_{\textrm{recipe}}(z,N) - \rho_{\textrm{target}}(z) \right]^2 \,dz$ to yield a set of $N$ $\{E_i, f_i\}$ pairs.

\subsection{Finding the number of implants to meet the error limit}
Finally, we find the minimum number of energies such that the percent error ($\Delta$) between $\rho_{\textrm{recipe}}(z,N)$ and $\rho_{\textrm{target}}(z)$ is less than a chosen threshold, where $\Delta$ (evaluated numerically) is defined as:
\begin{equation}
  \Delta =   \frac{ \int_{0}^{\infty} | \rho_{\textrm{recipe}}(z,N) - \rho_{\textrm{target}}(z) | ~dz}{ \int_{0}^{\infty} \rho_{\textrm{target}}(z) \,dz } \times 100\%  ~. 
\end{equation}
\noindent
Although increasing $N$ typically decreases $\Delta$, this also adds more implantation steps and cost. To avoid generating unnecessarily complicated ion implantation recipes, we find the smallest $N$ for which the $\Delta$ discrepancy is satisfactory. For a given $N$ we perform a least-squares fit for the set of $\{E_i, f_i\}$ energies and fluences, compare to a desired threshold $\Delta_{\textrm{th}}$ (for example, $\Delta_{\textrm{th}} = 10$\%), and repeat with increasing $N$ until we fulfill our $\Delta \leq \Delta_{\textrm{th}}$ requirement. 

\begin{table}[]
\begin{tabular}{c|c|c|c}
\begin{tabular}[c]{@{}c@{}}\nFifteen energy\\ (keV)\end{tabular} & \begin{tabular}[c]{@{}c@{}}\nFifteen fluence\\ (ions/cm$^2$)\end{tabular} & \begin{tabular}[c]{@{}c@{}}\heFour energy\\ (keV)\end{tabular} & \begin{tabular}[c]{@{}c@{}}\heFour fluence\\ (ions/cm$^2$)\end{tabular} \\
\hline
10      & 1.3E13 & 8      & 1.2E12 \\
25      & 2.0E13 & 23      & 1.4E12 \\
48     & 2.8E13 & 47      & 1.6E12 \\
79     & 3.4E13 & 80      & 1.8E12 \\
120     & 4.0E13 & 121     & 1.8E12 \\
170     & 4.8E13 & 170     & 1.9E12 \\
229     & 5.2E13 & 227     & 2.0E12 \\
299     & 5.9E13 & 288     & 2.0E12 \\
384    & 6.4E13 & 354     & 2.1E12 \\
483    & 6.6E13 & 422     & 2.2E12 \\
599    & 7.0E13 & 494     & 3.8E12 \\
734    & 7.4E13 &      & \\
891    & 7.7E13 &      &  \\
1064    & 7.6E13 &      &  \\
1255    & 7.9E13 &      &  \\
1478    & 9.8E13 &      &  \\
\end{tabular}
\caption{Calculated \nFifteen and \heFour implant energies and fluences for a 1 $\upmu$m 50 ppm defect layer, also shown in Fig.~\ref{exampleDoses}. }
\label{doseTable}
\end{table}

\section{Results and discussion}
In order to make an NV surface-layer diamond sample ideal for magnetically sensing external sources $\sim$1-10 $\upmu$m away from the surface, we want to create an implant recipe that has a uniform 50 ppm \nFifteen or vacancy density and a 1 $\upmu$m thickness.  We define this target defect density profile as:

\begin{equation}
  \rho_{\textrm{target}}(z) =
    \begin{cases}
      50 \text{~ppm} & \text{for 0 nm} \leq z \leq 1000 \text{~nm,}\\
      0 \text{~ppm} & \text{for 1000 nm} < z \leq 1300 \text{~nm.}\\
    \end{cases}
\end{equation}

\noindent
Figure \ref{exampleDoses}a-b shows the defect density profiles generated by applying the above algorithm.  We find that a minimum of sixteen \nFifteen implants and eleven \heFour implants are required to achieve $\Delta < 10\%$. Table \ref{doseTable} lists the energies and fluences for each fit in Fig.~\ref{exampleDoses}. For both \nFifteen and \heFour optimizations, $\Delta$ converges to about 3\%. This is because for shallow $z$, we become limited by the lineshape of the shallowest simulated implant. With increasing $N$, the fitting algorithm reduces $E_1$ (the energy of the lowest-energy implant) until $E_1$ equals the minimum available ion energy. In addition, the fit becomes overparameterized for sufficiently large $N$, which is realized when the fit yields duplicate energies or fluences that are approximately zero.

\section{Conclusions and outlook}
In this work, we described a way to empirically fit the simulated defect density profiles simulated by SRIM, predict the expected defect density profile for any energy within the simulated range, and calculate the ideal set of energies and fluences for a desired defect density profile (such as a uniform flat-top surface layer). By doing these steps computationally with least-squares fitting, we avoid having to guess the necessary implant parameters to  formulate the implant recipe. This method is generalizable to other situations, such as calculating non-uniform defect densities to compensate for depth-dependent NV conversion efficiency and coherence time \cite{creation_efficiency, schenkelSwiftHeavyIons, ania_surfNoise}, and also to creating defect layers with other ions and solids.  After implanting diamond samples with \nFifteen based on this algorithm and annealing to improve the NV yield, we were able to produce diamond samples with $\sim$1 $\upmu$m NV surface layers to use for NV magnetic microscopy applications \cite{QDM1ggg, edlynQDMreview}. The results of this work can be validated using techniques such as confocal microscopy or secondary-ion mass spectrometry (SIMS), and a future algorithm version can include defect diffusion during annealing.

\section{Acknowledgements}
We thank Heejun Byeon for help with critical review. Sandia National Laboratories is a multi-mission laboratory managed and operated by National Technology and Engineering Solutions of Sandia, LLC, a wholly owned subsidiary of Honeywell International, Inc., for the DOE's National Nuclear Security Administration under contract DE-NA0003525.  This work was funded, in part, by the Laboratory Directed Research and Development Program and performed, in part, at the Center for Integrated Nanotechnologies, an Office of Science User Facility operated for the U.S.~Department of Energy (DOE) Office of Science. This paper describes objective technical results and analysis. Any subjective views or opinions that might be expressed in the paper do not necessarily represent the views of the U.S. Department of Energy or the United States Government. P.K.~is supported by the Sandia National Laboratories Truman Fellowship Program. Example Python analysis code and SRIM simulation data are available for download \cite{suppl}.

\end{document}